\documentclass[preprint,12pt]{elsarticle}

\usepackage{amssymb}
\usepackage{amsmath}
\journal{}

\begin{document}

\begin{frontmatter}



\title{Quantifying Systematic Uncertainties in Experimental Physics: An Approximation Method}


\affiliation[inst1]{organization={Microsoft},%Department and Organization
            addressline={One Microsoft Way}, 
            city={Redmond},
            postcode={98052}, 
            state={WA},
            country={US}}
\author[inst1]{Lu Li}

\begin{abstract}
In the domain of physics experiments, data fitting is a pivotal technique for extracting insights from both experimental and simulated datasets. This article presents an approximation method designed to estimate the systematic errors prevalent in data analyses. By applying our method to the Nab experiment, we compare our findings with simulation-derived results, thereby confirming the concordance of our approach with established simulation outcomes. This corroboration highlights the versatility of our method as a good tool for validating simulation results across various experimental contexts.
\end{abstract}

\end{frontmatter}

\section{Introduction to Fitting}
Fitting is an essential technique in physics for revealing the underlying relationships in experimental or observational data, enabling the alignment of theoretical models with experimental outcomes. The primary goal of function fitting is to adjust a model's parameters, as predicted by scientific principles, to closely match observed data.

The least squares method is commonly employed in function fitting, aiming to minimize the sum of squared residuals between observed data values and their theoretical counterparts predicted by the model. The sum of squared errors (SSE) is mathematically formulated as\cite{Eadie:1971qcl}:

\begin{equation}
S = \sum_{i=1}^{N} \left[ y_i - f(x_i, \vec{\theta}) \right]^2,
\end{equation}
where \(y_i\) denotes the observed data at point \(i\), \(f(x_i, \vec{\theta})\) is the model's predicted value at \(i\) given parameters \(\vec{\theta}\), and \(N\) represents the total data points in the experiment. The optimization process involves selecting \(\vec{\theta}\) to minimize \(S\).

When measurements are associated with uncertainties, the SSE is adjusted by weighting each data point according to its uncertainty, leading to an optimized criterion\cite{Pedroni_2020, Cabrera_2012}:

\begin{equation}
S = \sum_{i=1}^{N} \left( \frac{y_i - f(x_i, \vec{\theta})}{\sigma_i} \right)^2,
\end{equation}
where \(\sigma_i\) is the measurement uncertainty of \(y_i\).

For linear models, the optimal \(\vec{\theta}\) can be directly determined. However, for more complex models, numerical methods like Monte-Carlo simulation are used to identify the best parameters\cite{Newman_1999}.
\section{Fitting Histograms with the Least Squares Method}

Histogram fitting represents a specialized form of function fitting, where the distribution of experimental data is compared against theoretical predictions. A histogram visualizes the distribution of numerical data by organizing data points into bins according to their values. The creation of a histogram from a dataset typically follows these steps:

\begin{enumerate}
    \item \textbf{Determination of bin edges:} The dataset's range is segmented into intervals or bins. The selection of bin size plays a crucial role in influencing the histogram's visual representation and the ensuing fitting process.
    \item \textbf{Allocation of data points to bins:} Data points are categorized into bins based on their values, ensuring each falls within the appropriate bin edges.
    \item \textbf{Counting the frequency:} The frequency of data points within each bin is counted, producing the histogram's y-values.
\end{enumerate}

The error associated with each histogram bin plays a critical role. Assuming the probability of a data point falling into a specific histogram bin \(i\) is \(p_i\), the distribution within bin \(i\) can be approximated by a Poisson distribution assuming each bin contains a large number of counts\cite{BAKER1984437}. The error \(\sigma_i\) for bin \(i\) with count \(y_i\) is expressed as:

\begin{equation}
\sigma_i = \sqrt{y_i}.
\end{equation}

Accounting for background noise—presumed independent from the primary events of interest—modifies the error for each bin, which varies based on the histogram's construction method. A typical approach to constructing the histogram includes:

\begin{enumerate}
    \item Measuring solely the background noise in the absence of the primary event of interest.
    \item Recording all events, both from the primary interest and background noise, over the same duration and categorizing them into histogram bins.
    \item Subtracting the background-only histogram (from step 1) from the combined event histogram (from step 2) to isolate the net value of each histogram bin.
\end{enumerate}

Employing this methodology, the error for each histogram bin is refined to:

\begin{equation}
\begin{split}
\sigma'_i &= \sqrt{(y_i + \sigma^2_\text{noise}) + \sigma^2_\text{noise}}\\
     &= \sqrt{y_i + 2\sigma^2_\text{noise}}.
\end{split}
\label{errors}
\end{equation}

Subsequent sections will utilize Eq.\,\eqref{errors} for error estimation across all histogram bins.

\section{Systematic Uncertainty and the Approximation Method}

Systematic uncertainty significantly impacts scientific experiments, setting itself apart from statistical uncertainty through its deterministic nature. Unlike statistical errors, which randomly affect measured data, systematic uncertainty arises from inherent flaws in the measurement process or experimental design, introducing a consistent bias in the data. This type of uncertainty can skew experimental results either above or below their true values and is not mitigated by increasing the size of the dataset. Therefore, understanding and quantifying the impact of systematic uncertainty on experimental outcomes is crucial.

To analyze the influence of systematic uncertainty, consider events characterized by a probability distribution \(p(x) = \Gamma(x, \vec{\theta_0})\), where \(\vec{\theta_0}\) represents the true, yet unknown, parameter value of \(\vec{\theta}\). Upon observing \(N_0\) events, the count \(y_i\) in the \(i\)th histogram bin is expected to follow a Poisson distribution with mean \(\mu_i = N_0\int_{{x'}^\text{min}_i}^{{x'}^\text{max}_i}\Gamma(x', \vec{\theta_0})dx'\)\cite{BOHM2012171}. Here, \(x'\) denotes the detected value of \(x\), which, in the presence of perfect detection, would equal \(x\). However, systematic uncertainty in the detection system means \(x'\) is a function of \(x\), with the nature of systematic uncertainty determining the form of \(x'(x)\).

Given this framework, the fitting equation becomes:
\begin{equation}
S = \sum_{i=1}^{K} \left[\frac{\mu_i - N\Gamma(x^{\text{mid}}_i, \vec{\theta})}{\sigma'_i}\right]^2.
\end{equation}
In this context, \(x^\text{mid}_i\) represents the midpoint of the histogram bin. In practical scenarios, such as neutron beta decay experiments, the exact number of decay events (\(N_0\)) may be ambiguous due to background noise. Consequently, fitting often treats the total number of decay events as an adjustable parameter.

A more refined approximation of the fitting equation is possible when the histogram is sufficiently detailed. In this scenario, the probability distribution is almost constant within each histogram bin, allowing the summation to be approximated by integration:
\begin{equation}
S = \int_{x^\text{lower}}^{x^\text{upper}} \left[\frac{N_0\Gamma(x'(x), \vec{\theta_0})\frac{dx'}{dx} + \epsilon(x') - N\Gamma(x, \vec{\theta})}{\sigma'(x')}\right]^2dx.\footnote{Here, \(x^\text{lower}\) and \(x^\text{upper}\) define the histogram's fitting range, $\epsilon(x')$ is the statistical error associated with the Poisson distribution, which will be ignored since we are only interested in the effect of systematic uncertainties.}
\end{equation}

Without systematic uncertainties, the optimal parameter selection that minimizes \(S\) would simply equate \(N\) to \(N_0\) and \(\vec{\theta}\) to \(\vec{\theta_0}\). However, the presence of systematic uncertainty means the optimal values of \(N\) and \(\vec{\theta}\) depend on \(x'(x)\), illustrating the significant impact of systematic uncertainty on experimental findings.
\section{Analyzing Systematic Uncertainties in the Nab Experiment}

In the pursuit of precision in experimental physics, understanding and quantifying systematic uncertainties is paramount. This section applies our method to assess systematic uncertainty propagation within the context of the Nab experiment. One of the goal of the Nab experiment is to measure the Fierz interference term \(b\) using the electron energy spectrum from free neutron beta decay\cite{fry2019nab, broussard2017neutron}.

The decay rate of electron in this context is modeled as\cite{PhysRev.106.517}:
\begin{equation}
\Gamma(E_e, b) = \frac{F(Z,E_{e})}{{(2\pi)}^{5}}  p_{e}E_{e}(E_{0}-E_{e})^{2}(1+b\frac{m_e}{E_e + m_e}),
\end{equation}
where \(E_e\) represents the kinetic energy of the electron, and \(m_e\) is the electron mass. The experiment employs highly linear silicon detectors for capturing outgoing protons and electrons \cite{broussard2017detection, 1968NucIM..59...45P}. Achieving the desired experimental precision necessitates meticulous calibration of the detector's linearity. Studies by \cite{Li:2021oal, li2016systematic} have explored the systematic uncertainties arising from detector miscalibration, particularly focusing on gain factor and offset, described by:
\begin{equation}
E_e' = g E_e + o.
\end{equation}

Incorporating this into our systematic uncertainty analysis yields:
\begin{equation}
S = \int_{E_e^\text{lower}}^{E_e^\text{upper}} \left[\frac{gN_0\Gamma(g E_e + o, b_0) - N\Gamma(E_e, b)}{\sqrt{gN_0\Gamma(g E_e + o, b_0) + 2\sigma^2_\text{noise}}}\right]^2dE_e.
\end{equation}

\begin{figure}[h!]
\centering
\includegraphics[scale=0.25]{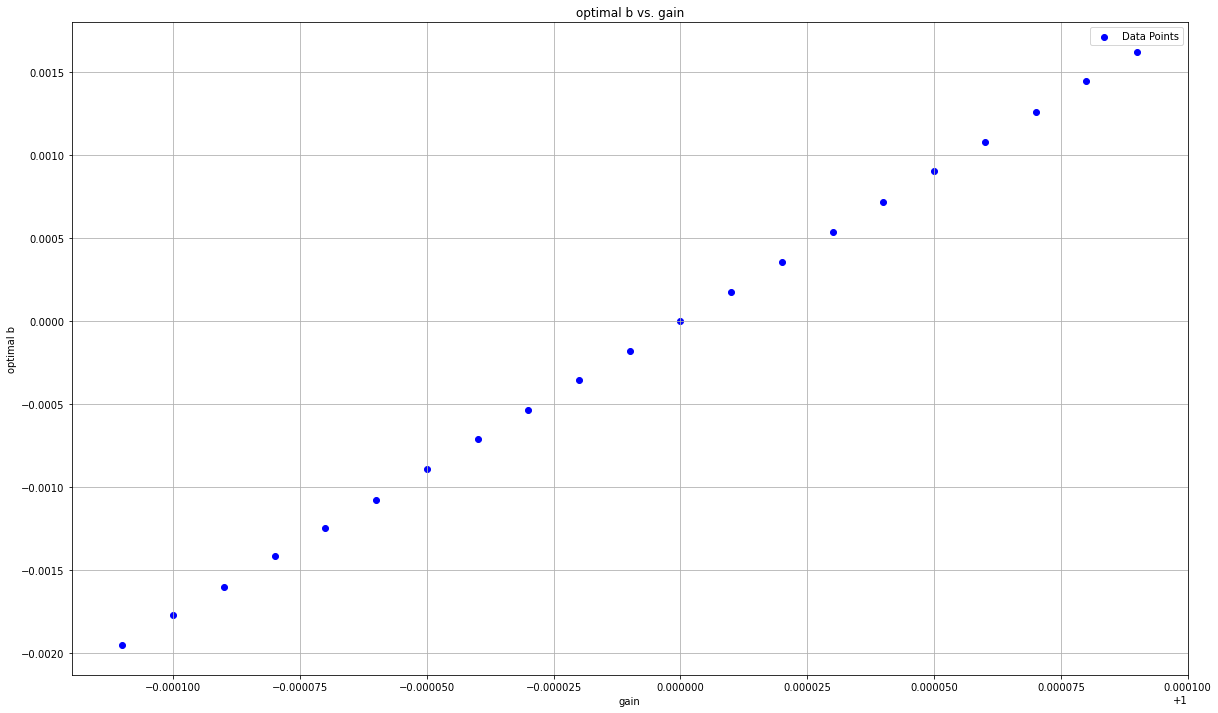}
\caption{The impact of gain factor miscalibration on the optimal value of \(b\).}
\label{g1}
\end{figure}

\begin{figure}[h!]
\centering
\includegraphics[scale=0.25]{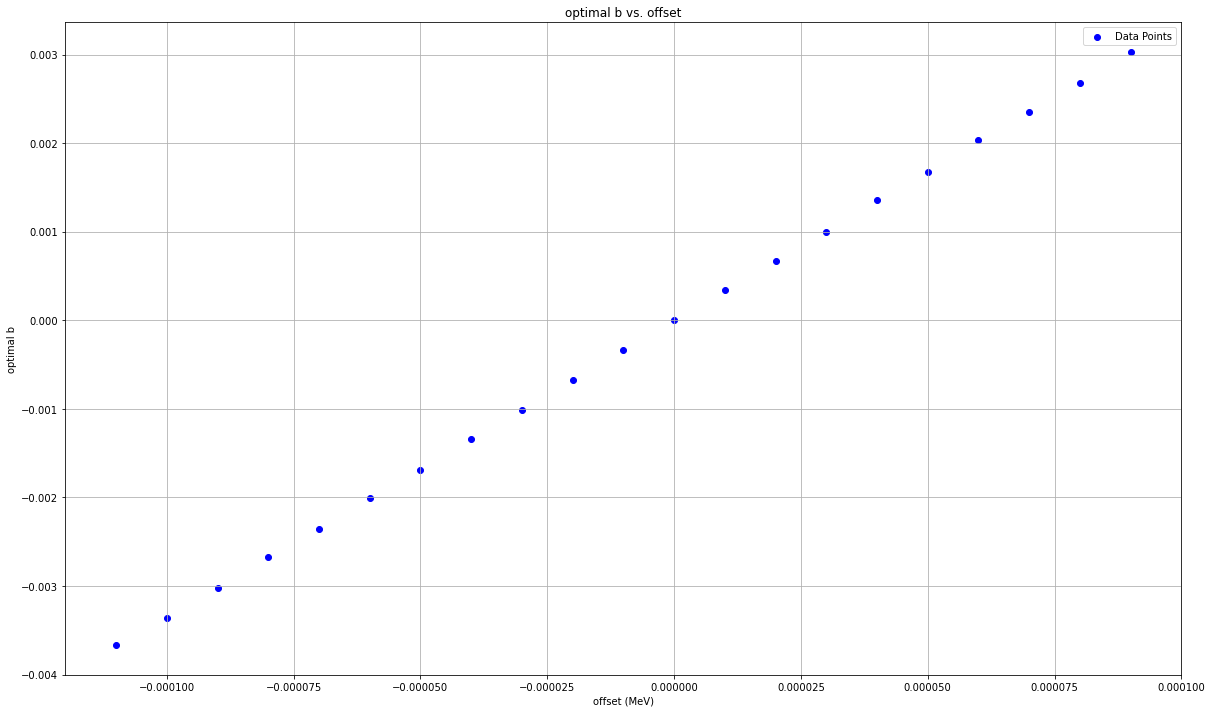}
\caption{The influence of offset miscalibration on the optimal value of \(b\), with a fixed gain value.}
\label{g2}
\end{figure}

\begin{figure}[h!]
\centering
\includegraphics[scale=0.25]{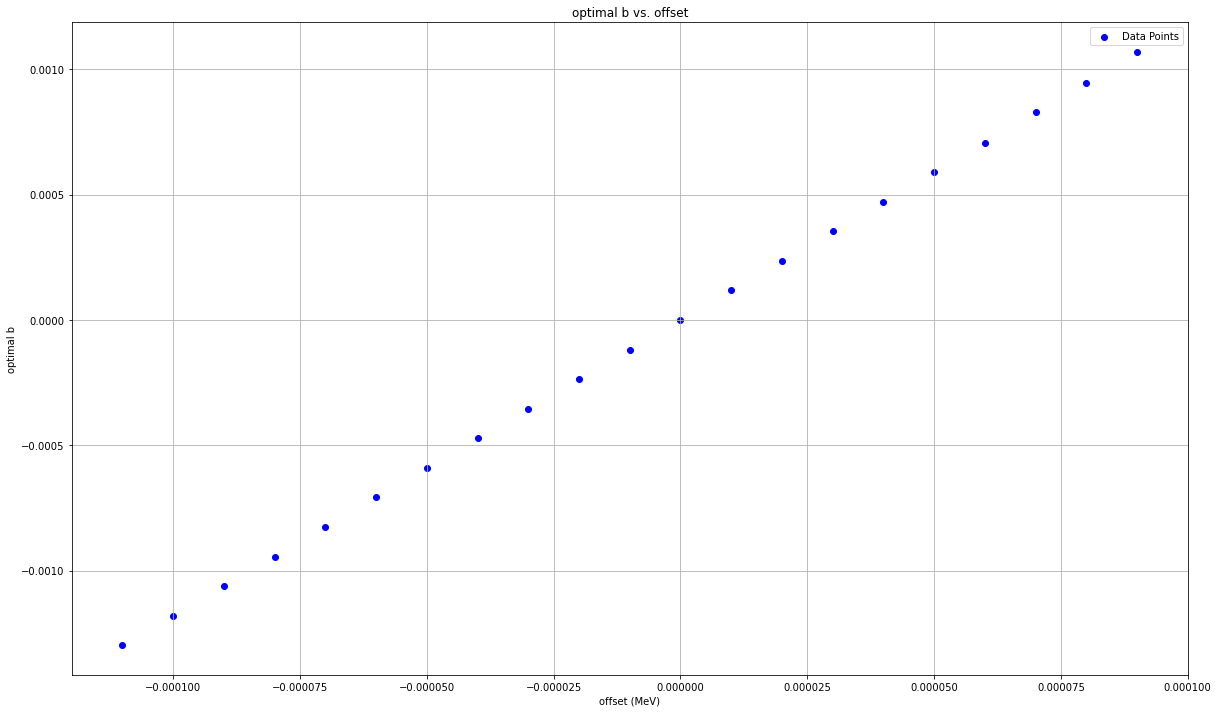}
\caption{The effect of offset miscalibration on the optimal value of \(b\), with gain as a fitting parameter.}
\label{g3}
\end{figure}

The fitting range spans from 0.16 MeV to 0.78 MeV, with the simulation error comprising both the histogram height uncertainty and twice the background noise uncertainty (10\% of decay events), setting \(b_0\) to 0. This leads to:
\begin{equation}
S_w = N_0\int_{0.16}^{0.78} \left[\frac{(g\Gamma(g E_e + o, 0) - r\Gamma(E_e, b))^2}{g\Gamma(g E_e + o, 0)+ 0.2}\right]dE_e,
\end{equation}
where \(r = \frac{N}{N_0}\) is considered a fitting parameter. Through numerical integration, we explored various values of \(g\) and \(o\), extracting the corresponding \(b\) values. Figures \ref{g1} to \ref{g3} illustrate how the optimal \(b\) value is influenced by \(g\) and \(o\).

To synthesize our findings, we compared our results with Monte-Carlo simulation outcomes from \cite{Li:2021oal}, as summarized in Table \ref{systable}.

\begin{table}[h!]
\begin{center}
\begin{tabular}{|l|c|c|}
\hline
Parameter & Slope (Simulation) & Slope (Calculation) \\
\hline
Gain factor & 17.9 & 17.9 \\
Offset (fixed gain) & 33.7/MeV & 33.6/MeV \\
Offset (variable gain) & 11.8/MeV & 11.8/MeV \\
\hline
\end{tabular}
\caption{Comparison of systematic uncertainty effects using Monte-Carlo simulations and our approximation calculations.}
\label{systable}
\end{center}
\end{table}

We find that our result agrees well with the Monte-Carlo simulation results. Unlike the Monte-Carlo simulation results, our calculation doesn't make assumptions about the binning of the histogram, and is more general on the relation between the systematic uncertainty parameters and the extracted optimal $b$ values.

\section{Conclusion}

In this article, we introduced an approximation technique designed to estimate systematic errors within data analyses, with a particular focus on the Nab experiment. The alignment of our method with simulation results, alongside its proficiency in quantifying the effects of systematic uncertainties—predominantly arising from detector calibration errors—underscores its utility and importance. Our methodology serves as a robust tool for validating simulation outcomes and broadening the scope of analysis to encompass more intricate systematic uncertainties.

\appendix

 \bibliographystyle{elsarticle-num} 
 \bibliography{cas-refs}





% \end{thebibliography}
\end{document}